\documentclass[twocolumn,showpacs,amssymb,aps, prl]{revtex4}
\usepackage{tipa}
\usepackage{graphicx}
\usepackage{dcolumn}
\usepackage{bm}
\usepackage{amsmath}
\usepackage{amssymb}
\usepackage{natbib}
\usepackage{epsfig}
\begin{document}

\title{Electron Temperature Dependence of the Optical Properties of\\ Small Sodium Nanoparticles}
\author{Guozhong~Wang$^1$}
\email{gzw@wzu.edu.cn;jzi@fudan.edu.cn}
\author{Yizhuang~Zheng$^1$}
\author{Jian~Zi$^2$$^\dag$}
\affiliation{$^1$School of physics and electric information, Wenzhou
University, Wenzhou 325035, P.R.China\\$^2$ Department of Physics,
Fudan University, Shanghai 200433, P.R.China}
\date{\today}

\begin{abstract}
We report a novel behavior of the surface plasmon linewidth in
sodium nanospheres Na$_{1760}$ changing with the electron
temperature, which monotonically decreases and bears a discontinuous
sudden drop at high electron temperatures. Our calculation is based
on the model constructed by splitting the total Hamiltonian of all
valence electrons of a metallic nanoparticle into two
sub-Hamiltonians and their coupling, and obtained results can be
verified by the pump-probe femtosecond spectroscopy experiments. In
addition, we propose that it is the size uncertainty of small
nanoparticles that yields the intrinsic linewidth of the surface
plasmon resonance, which is supported by the available data of
clusters Na$_{8}$ and Na$_{20}$.
\end{abstract}
 \maketitle

When illuminated by a light with proper frequency, the delocalized
valence electrons of a metallic nanoparticle (MNP) will be excited
to perform collective oscillations against ions, the so-called
surface plasmon resonance (SPR). MNPs could enable strong optical
absorption and scattering in deep subwavelength scale, with spectral
properties determined by their composition, size, geometry, charge
and surrounding medium \cite{ukr}. The application based on
different features of SPR flourished in many areas, such as
surface-enhanced Raman scattering \cite{pau}, single molecule
detection \cite{pet}, plasmon rulers \cite{nal}. Besides
nanophotonics, SPR is also finding use in biology, catalysis,
quantum optics and quantum information transfer. The capability of
plasmonic systems to manipulate and enhance optical signals at
nanoscale has paved the way for many novel concepts and
applications, which requires a deep understanding of physics
occurred in MNPs.

Based on the features of MNPs interacting with lights, different
models and theories have been constructed. For large MNPs
($\ge100$\,nm), the quantum effect is negligible and their optical
properties can be well described by classical Drude dielectric
function $\epsilon(\omega)=1-\omega^2_p/\omega(\omega+i\gamma_i)$,
where $\omega_{p}=(4\pi\rho_e e^2/m)^{1/2}$ is the Mie frequency of
the plasmon resonance, $\gamma_i^{-1}$ the relaxation time, and $e$,
$m$ and $\rho_e$ the electron charge, mass and density respectively.
Classical plasmonic systems yield resonance at Mie frequency which
does not depend on MNPs' size. For quantum plasmonic systems, the
time-dependent local density approximation (TDLDA) and all kinds of
random phase approximation methods (RPA) are powerful tools to study
their optical properties \cite{mat,wek}. For medium-sized MNPs with
radii $10-60$\,nm, their optical properties are quite difficult to
accurately describe as the situations of very small and very large
MNPs. Many non-fundamental methods were also developed for the
optical properties of MNPs with sizes from several nanometers to
several tens of nanometers \cite{jsa,edu,jon,rub}. For MNPs smaller
than $5$\,nm, the model constructed by separating the coordinates of
all delocalized valence electrons into the coordinate of the center
of mass and relative coordinates describes the SPR as the
oscillation of the center-of-mass, and the SPR is damped by the
coupling between the center of mass of valence electrons and
electrons outside the nanoparticle (SCRM) \cite{gui,lgg,raf}.

Although the SCRM is fundamental and elegant, the results done
previously are not satisfactory. As a matter of fact, the best
founded TDLDA and all kinds of RPA methods are not perfect either.
It is believed that the full linewidth of the SPR comprises an
intrinsic part $\gamma_i$ and a size-dependent part $\gamma_s(a)$.
However, neither TDLDA nor RPA methods touch upon the calculation of
the intrinsic linewidth, and its generation mechanism is still an
open question. For bulk metals, the intrinsic linewidth of the SPR
is caused by the scattering of delocalized electrons by phonons,
impurities and defects. For small MNPs, electron states are
quantized into discrete levels due to strong quantum confinement
effect and the scattering of a delocalized electron experienced in
bulk metals does not exist, which foreshadows a different generation
mechanism for the intrinsic linewidth of small MNPs. In this letter,
we will use SCRM to study the optical properties of nanospheres
Na$_{1760}$ in detail and explore the generation mechanism of the
intrinsic linewidth.

For alkali metals, the optical properties is mainly determined by
delocalized valence electrons, and ionic cores can be treated as a
uniform background \cite{mat}. For nanoparticles with valence
electrons closing angular momentum shells, such as Na$_{1760}$,
their shapes can be thought of as perfect spheres. For such a sodium
nanoparticle composed of $N$ atoms in vacuum, by introducing the
coordinate of the center of mass $\bold{R}=(\sum{\vec{\,r}_i})/N$
and its conjugate momentum $\bold{P}=\sum\vec{p}_i$, the total
Hamiltonian $\mathcal{H}$ of all delocalized valence electrons
denoted by coordinate and momentum pairs $(\vec{\,r}_i,\,\vec{p}_i)$
can be separated into three parts $\mathcal H=\mathcal H_1+\mathcal
H_2+\mathcal H_{12}$, and the detailed derivation can be found in
the supplementary information. Three terms of total Hamiltonian
$\mathcal H$ are
\begin{eqnarray}
&&\qquad\qquad \mathcal {H}_1=\frac{\bold{P}^2}{2Nm}+\frac{1}{2}Nm\Omega^2_p\bold{R}^2\nonumber\\
&&\mathcal {H}_2=\sum_{i=1}^{N}(\frac{\vec\varrho_i^{\,2}}{2\,m}+U(
\xi_i))+\frac{e^2}{8\pi\varepsilon_0}\sum_{i\neq j}^{N}\frac{1}{|\vec{\,\xi}_i-\vec{\,\xi}_j|}\nonumber\\
&&\,\mathcal
{H}_{12}=\frac{Ne^2}{4\pi\varepsilon_0a^3}\bold{R}\cdot\sum_{i=1}^N\vec{\,\xi}_i(\frac{a^3}{\xi_i^3}-1)\Theta(\xi_i-a)
\end{eqnarray}
respectively, where $\xi_i=|\hspace{-1pt}\vec{\,\xi_i}|$; $a$ and
$\Theta(x)$ are the radius of nanospheres and the Heaviside step
function; $\vec{\,\xi}_i=\vec{\,r}_i-\bold{R}$ and
$\vec{\varrho}_i=\vec{\,p}_i-\bold{P}/N$ are the relative coordinate
and relative momentum of the $i$-th electron. The single electron
confining potential $U(|\vec\xi+\bold{R}|)$ produced by ionic
background was expanded to the second order at $|\bold R|=0$. The
Hamiltonian $\mathcal {H}_1$ has the harmonic oscillator structure
with the frequency $\Omega_p=\omega_p\sqrt{1-N_{out}/N}$ describing
the motion of the center of mass, where
$N_{out}=\sum_i^N\Theta(\xi_i-a)$ is the number of spill-out
electrons. The quanta of the Hamiltonian $\mathcal{H}_1$ correspond
to SPR excitations in MNPs. The Hamiltonian $\mathcal{H}_2$
describes the degrees of freedom of relative coordinates of all
valence electrons. The third part $\mathcal{H}_{12}$ shows that the
SPR couples not to electrons inside the nanosphere but to those
outside the nanosphere.

The Hamiltonian $\mathcal {H}_1$ can be further expressed as
$\mathcal {H}_1=\sum(n+1/2)\hbar\Omega_p\,\hat{b}^{\dag}\hat{b}$,
and the annihilation operator $\hat{b}$ is defined as
\begin{equation}\label{eq2}
\hat{b}=\sqrt{\frac{Nm\Omega_p}{2\hbar}}\bold{R}+\frac{i}{\sqrt{2Nm\hbar\Omega_p}}\bold{P},
\end{equation}
where $i$ is the imaginary unit. The potential energy part of
$\mathcal{H}_2$ is usually approximated by a mean field which
transforms the many body problem into a single particle problem, so
the $\mathcal{H}_2$ can be expressed as
$\mathcal{H}_2=\sum_\alpha\epsilon_\alpha\,\hat{c}^{\dag}_\alpha
\hat{c}_\alpha$. The energy levels and corresponding wavefunctions
of $\mathcal{H}_2$ can be obtained by solving the Schr\"{o}dinger
equation
\begin{equation}\label{eq3}
(-\frac{\hbar^2}{2m}\nabla^2+V_{eff}(\xi))\psi_{\alpha}(\vec{\,\xi})=\epsilon_{\alpha}\psi(\vec{\,\xi}),
\end{equation}
where the single particle effective potential $V_{eff}$ can be
obtained by LDA calculation \cite{gui}. Because a systematic
multi-step approximation method has been developed by present
authors to solve Schr\"odinger equation with a Woods-Saxon-like
potential \cite {guo}, total $134$ energy levels and corresponding
piecewise analytical wavefunctions with high accuracy are obtained
for nanospheres Na$_{1760}$, which makes it possible to accurately
calculate the optical properties of Na$_{1760}$.

\begin{figure}
\vspace*{5pt}
\includegraphics*[width=0.4\textwidth]{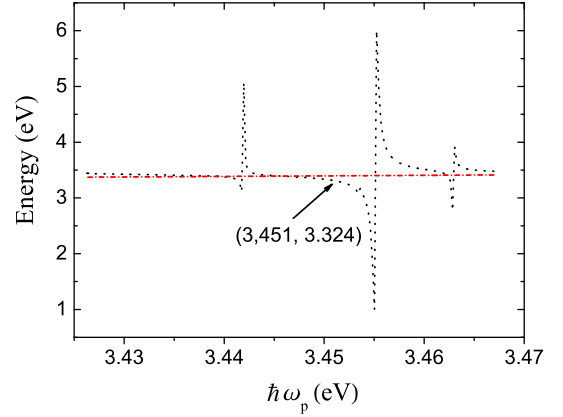}
\vspace{-10pt} \caption{The SPR energy $\hbar\Omega_q$ (dotted line)
calculated according to Eq. (\ref{eq6}) and $\hbar\Omega_p$ (red
line) for Na$_{1760}$ versus the classical plasmon energy
$\hbar\omega_p$. The divergent values corresponding to
$\epsilon_{\alpha\beta}\sim\Omega_p$ have been properly minished for
the sake of display.} \label{fig:1} \vspace*{-15pt}
\end{figure}

The coupling $\mathcal {H}_{12}$ can be recast as
\begin{equation}\label{eq4}
\mathcal
{H}_{12}=\mathcal{A}(\hat{b}^{\dag}+\hat{b})\sum_{\alpha,\beta}d_{\alpha\beta}\hat{c}^{\dag}_\alpha
\hat{c}_\beta,
\end{equation}
where the coefficient $\mathcal
{A}=\frac{e^2}{4\pi\varepsilon_0a^3}\sqrt{\frac{\hbar
N}{2m\Omega_p}}$, the matrix element $d_{\alpha\beta}=\langle
\alpha|\xi_z(a^3/\xi^3-1)\Theta(\xi-a)|\beta\rangle$ is calculated
between two states $|\alpha\rangle$ and $|\beta\rangle$ of $\mathcal
{H}_{2}$.

Although the frequency $\Omega_p$ is always smaller than its
classical counterpart $\omega_p$, the redshift resulted just by the
spill-out effect of valence electrons is not enough to explain
experimental observations. The real frequency of SPR is the
difference between the ground state and the first excited state of
$\mathcal {H}_1$ with the second order perturbation corrections
produced by the coupling $\mathcal{H}_{12}$, namely
$\hbar\Omega_q=\hbar(\Omega_p-\Delta)$ and
\begin{equation}\label{eq5}
\Delta=\sum_{\alpha\beta}2\mathcal{A}^2[1-f(\epsilon_\alpha)]f(\epsilon_\beta)
\frac{|d_{\alpha\beta}|^2\epsilon_{\alpha\beta}}{\epsilon_{\alpha\beta}^2-(\hbar\Omega_p)^2},
\vspace{-6pt}
\end{equation}
where $\epsilon_{\alpha\beta}=\epsilon_\alpha-\epsilon_\beta$, and
$f(\epsilon)=1/(1+e^{(\epsilon-\mu)/k_{B}T})$ is the Fermi-Dirac
distribution; $T$, $k_B$ and $\mu$ the electron temperature, the
Boltzmann constant and the chemical potential respectively. The
values of $\mathcal{A}$ and $\Omega_p$ are determined by the radius
$a=N^{1/3}r_s$, and $r_s$ is the Wigner-Seitz radius. However, the
value of $r_s$ for sodium nanoparticles appeared in references
varies from $3.93a_0$ to $4.3a_0$ being quite dispersive
\cite{mat,wek,gui,con}, where $a_0=0.529$\,\r{A} is the Bohr radius.
We calculated the energy of the SPR of Na$_{1760}$ changing with
$\hbar\,\omega_p$ and results are shown in Fig.~\ref{fig:1}. Only
within a very narrow range of $\hbar\,\omega_p$ does the coupling
$\mathcal{H}_{12}$ yield the frequency redshift of the SPR. The SPR
energy $\hbar\Omega_q=3.324$eV of the TDLDA calculation just locates
in this redshift range \cite{gui}, which corresponds to the radius
$a=25.313$\,\r{A} and $r_s=3.962a_0$. The cusps in Fig.~\ref{fig:1}
correspond to the situation that the Eq.~(\ref{eq5}) is invalid due
to the approximate degeneracy between the SPR state and
electron-hole states.

\begin{figure}
\vspace{3pt}
\includegraphics*[width=0.40\textwidth]{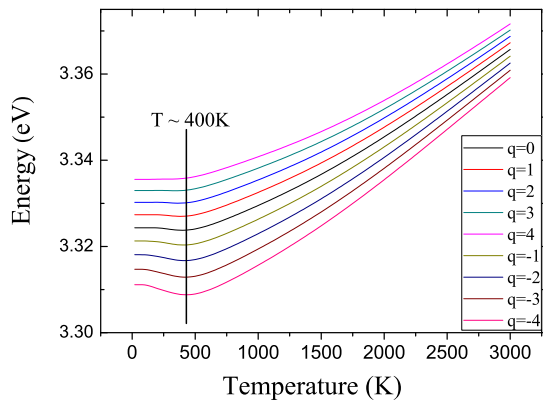}
\vspace{-5pt} \caption{The SPR energy of nanospheres Na$_{1760}$
with different charges as a function of the electron temperature.
The lowest line is for the situation of $q=-4|\,e|$, and $e$ is the
charge of an electron.} \label{fig:2} \vspace{-15pt}
\end{figure}

The evolution of the SPR energy of Na$_{1760}$ with different
charges with the electron temperature is presented in
Fig.~\ref{fig:2}, which first shows a redshift trend from zero
temperature up to $T\sim400$\,K and then presents a explicit
blueshift behavior. This evolution pattern is the same as those of
Na$_{138}$ and Na$_{139}^+$ \cite{pah}. However, it is astonishing
that the turning temperature ($1000$K for Na$_{138}$ and $2500$K for
Na$_{139}^+$) drastically changes with the charge of nanospheres,
which is quite different from our results.

The linewidth is another significant quantity to characterize the
optical properties of MNPs. The size-dependent linewidth of the SPR
in small MNPs is determined by Landau damping mechanism and can be
expressed as
\begin{equation}\label{eq6}
\hbar\gamma_s(a)=\mathcal
{B}\sum_{\alpha\beta}[1-f(\epsilon_\alpha)]f(\epsilon_\beta)|d_{\alpha\beta}|^{\,2}\delta(\epsilon_{\alpha\beta}-\hbar\Omega_p),
\vspace{-8pt}
\end{equation}
where $\mathcal{B}=\frac{\hbar \, e^4N}{16\pi
m\varepsilon_0^2a^6\Omega_p}$, and Dirac $\delta$ function
$\delta(\epsilon_{\alpha\beta}-\hbar\Omega_p)$ represents the
condition of energy conservation. However, there are no
particle-hole pairs of Na$_{1760}$ strictly satisfying
$\epsilon_{\alpha\beta}=\hbar\Omega_p$. By first writing
$\gamma_s(a)=\sum\gamma_{\alpha\beta}(a)$ and assuming that each
$\gamma_{\alpha\beta}(a)$ is the smooth function of its variables,
we then integrate Eq.~(\ref{eq6}) at both sides in a narrow interval
$[\hbar\Omega_p-\delta\epsilon, \hbar\Omega_p+\delta\epsilon]$,
where $\delta\epsilon$ is left to be determined. So all the $\delta$
functions of Eq.~(\ref{eq6}) can be replaced by $1/2\delta\epsilon$.
To single out the particle-hole pairs involved in the damping of the
SPR, each energy level should be broadened out according to a
generally used formula \cite{hni},
\begin{equation}\label{eq7}
\mathcal{E}(\epsilon,\epsilon_\alpha)=\frac{2}{\pi}\frac{\sqrt{\epsilon_{_T}\epsilon_\alpha}}{(\epsilon-\epsilon_\alpha)^2+4\epsilon_{_T}\epsilon_\alpha},
\end{equation}
where $\epsilon_{_T}=(\hbar k_0)^2/2m$, and
$k_0=0.13N^{-1/3}$\r{A}$^{-1}$. The width of a energy level changes
from zero to $2\sqrt{4\epsilon_{_T}\epsilon_\alpha}$ and the
condition of energy conservation of Eq. (\ref{eq6}) should be
replaced by
\begin{equation}\label{eq8}
|\hbar\Omega_p-(\epsilon_\alpha+\epsilon_\beta)|\le(\sqrt{4\epsilon_{_T}\epsilon_\alpha}+\sqrt{4\epsilon_{_T}\epsilon_\beta}).
\end{equation}
Therefore, a reasonable choice of $2\delta\epsilon$ is the smallest
average value of two energy widths of particle-hole pairs satisfying
Eq.~(\ref{eq8}), namely
$2\delta\epsilon=$\,Min\{$\sqrt{4\epsilon_{_T}\epsilon_\alpha}+\sqrt{4\epsilon_{_T}\epsilon_\beta}$\,\}.
Fig.~{\ref{fig:3} shows the size-dependent linewidth
$\hbar\gamma_s(a)$ of the SPR in Na$_{1760}$ with different charges,
which monotonically decreases with the increasing electron
temperature and bears a discontinuous sudden drop at high electron
temperatures. To our knowledge, this novel behavior of the SPR
linewidth has not been reported before, which is completely
different from those obtained by other authors. Weick et~al.
obtained the opposite result that the SPR linewidth of Na$_{1760}$
monotonically increases with the electron temperature \cite{gui}.
Hervieux and Bigot also obtained the result that the SPR linewidths
of Na$_{138}$ and Na$_{139}^+$ increase with the electron
temperature \cite{pah}. We notice that although the temperature
effect is included, the SPR linewidth of Na$_{138}$ calculated by
Hervieux and Bigot is quite smaller than that of matrix RPA
calculations \cite{con}. Our calculated result, the SPR
size-dependent linewidth ($0.224$eV) at the zero electron
temperature, is the same as that of TDLDA calculation \cite{gui}.
This novel behavior of SPR linewidth of Na$_{1760}$ can be easily
understood. When the electron temperature increases, valence
electrons have larger possibility to occupy the higher energy levels
leading to the smooth decrease of $\gamma_s(a)$. At the same time,
valence electrons have larger possibility to stay outside leading to
the slight decrease of the frequency $\Omega_p$, which in turn
alters the number of particle-hole pairs involved in the SPR damping
yielding the sudden drop of the linewidth. The sudden drop in
Fig.~{\ref{fig:3} is caused by the number of particle-hole pairs
decreasing from $29$ to $28$. In the following we will show that
this novel behavior of the SPR linewidth of Na$_{1760}$ would be
slightly enhanced by its intrinsic linewidth and could be verified
by the pumb-probe femtosecond spectroscopy experiments.

\begin{figure}
\vspace{4pt}
\includegraphics*[width=0.40\textwidth]{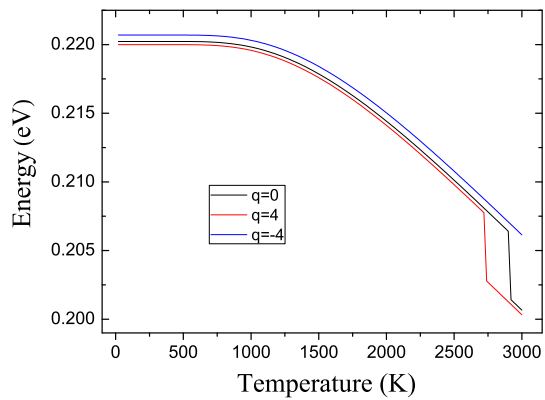}
\vspace{-6pt} \caption{The size-dependent linewidth
$\hbar\gamma_s(a)$ of nanoparticles Na$_{1760}$ with different
charges as a function of the electron temperature.} \label{fig:3}
\vspace{-17pt}
\end{figure}

The studies on the intrinsic linewidth of the SPR are precious few.
It is believed that the coupling of the dipole SPR to the quadrupole
shape fluctuations of positive ions produces the intrinsic linewidth
proportional to $(T/N)^{1/2}$ \cite{jmp}. However, this thermal
mechanism of the intrinsic linewidth is questionable, which would
lead to the corollary that the intrinsic linewidth vanishes when the
ions temperature decreases to zero. Furthermore, the lifetime of SPR
is too transient to effectively couple the SPR with the shape
fluctuations of ions. Due to the surface diffuseness and density
fluctuations of ions, there always exists a size uncertainty of a
MNP, which could yield an extra linewidth of the SPR. We propose
that this extra linewidth induced by the size uncertainty of MNPs is
the so-called intrinsic linewidth which can be expressed as
$\gamma_i=|\gamma_s(a+\delta a)-\gamma_s(a)|$. When the radius $a$
changes to $a+\delta a$, the energy level $\epsilon_\alpha$ and the
matrix element $d_{\alpha\beta}$ will change to
$\epsilon_\alpha+\delta\epsilon_\alpha$ and $d_{\alpha\beta}+\delta
d_{\alpha\beta}$ respectively. The quantum perturbation theory shows
that $\delta\epsilon_\alpha\propto|\psi_\alpha|^2\delta{a}/a$ and
$\delta d_{\alpha\beta}\propto\psi_\alpha^*\psi_\beta\delta{a}/a$.
Therefore the contributions to the intrinsic linewidth
$\hbar\gamma_i$ from variations $\delta\epsilon_\alpha$ and $\delta
d_{\alpha\beta}$ are negligible due to the fact that $|\psi|\ll1$.
The final expression of intrinsic linewidth is \vspace*{-5pt}
\begin{equation}\vspace*{-5pt}
\frac{\gamma_i}{\gamma_s(a)}=[1-\frac{1}{(1+\delta{a}/a)^{7.5}}],\label{eq9}
\end{equation}
where the size uncertainty $\delta{a}$ at zero temperature is about
$0.5\,a_0$. For nanoparticles Na$_{1760}$, the intrinsic linewidth
$\hbar\gamma_i$ is about $0.016$eV. The valuable data of Na$_8$ and
Na$_{20}$ support our proposal for the intrinsic linewidth, which
are depicted in Fig.~{\ref{fig:4}. Further details can be found in
the supplementary information. Eq.~(\ref{eq9}) shows that the
intrinsic linewidth would not weaken the novel behavior of the
size-dependetn linewidth of Na$_{1760}$. Although the
electron-phonon scattering could cause an apparent broadening of the
SPR, the different time scales of the SPR lifetime ($\sim$fs) and
energy exchange between valence electrons and the lattice ($\sim$ps)
guarantee that the pump-probe femtosecond spectroscopy techniques is
able to verify this novel behavior of the SPR linewidth of
Na$_{1760}$. In the final analysis, the size uncertainty mechanism
of the intrinsic linewidth is the manifestation of Landau damping
mechanism in a MNP system with a diffusive surface. When MNPs become
larger, the intrinsic linewidth originating from the size
uncertainty will vanish and be superseded by that of bulk metals.

\begin{figure}
\vspace{4pt}
\includegraphics*[width=0.40\textwidth]{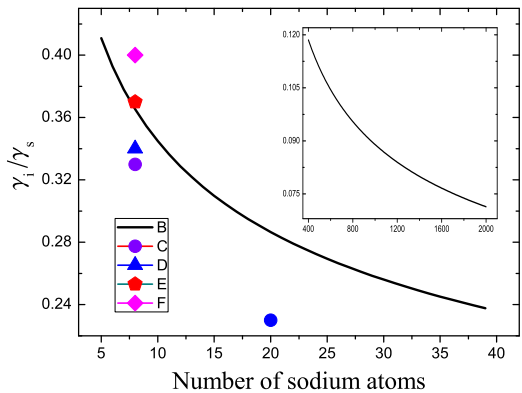}
\vspace{-6pt} \caption{The ratio between the intrinsic linewidth and
the size-dependent linewidth. The black line is calculated according
to Eq. (\ref{eq9}). The four symbols on the left side are the
experimental results for clusters Na$_8$. The blue solid circle is
for clusters Na$_{20}$.} \label{fig:4} \vspace{-17pt}
\end{figure}

In summary, we have studied the optical properties of Na$_{1760}$
and found a novel behavior of the linewidth for the first time,
which monotonically decreases with the electron temperature and
bears a discontinuous sudden drop at high temperatures. We think
that this novel behavior of the linewidth also exists in other
sodium nanospheres with closed angular momentum shells, but the
sudden drop maybe becomes sudden rise. We also propose that the
intrinsic linewidth of the SPR originates from its size uncertainty,
which essentially manifests the Landau damping mechanism in a system
without the exact size due to the surface diffuseness. The intrinsic
linewidth would enhance the novel behavior of the size-dependent
linewidth of Na$_{1760}$, and we expect that this novel behavior
could be verified in future by the elaborate pump-probe femtosecond
spectroscopy experiments.

\begin{acknowledgments}
This work was supported by the NSF of People's Republic of China
under contract No.10904115.
\end{acknowledgments}

\end{document}